# Characteristics of hand and machine-assigned scores to college students' answers to open-ended tasks

## Stephen P. Klein[*1]

*GANSK & Associates*

**Abstract:** Assessment of learning in higher education is a critical concern to policy makers, educators, parents, and students. And, doing so appropriately is likely to require including constructed response tests in the assessment system. We examined whether scoring costs and other concerns with using open-end measures on a large scale (e.g., turnaround time and inter-reader consistency) could be addressed by machine grading the answers. Analyses with 1359 students from 14 colleges found that two human readers agreed highly with each other in the scores they assigned to the answers to three types of open-ended questions. These reader assigned scores also agreed highly with those assigned by a computer. The correlations of the machine-assigned scores with SAT scores, college grades, and other measures were comparable to the correlations of these variables with the hand-assigned scores. Machine scoring did not widen differences in mean scores between racial/ethnic or gender groups. Our findings demonstrated that machine scoring can facilitate the use of open-ended questions in large-scale testing programs by providing a fast, accurate, and economical way to grade responses.

Until the turn of the 21st century, most large-scale K-12 and college testing programs relied almost exclusively on multiple-choice tests. There are several understandable reasons for this. It takes much longer to score the answers to essay and other "constructed response" (which are also referred to below as "open-ended" or "free-response") questions than it does to have a machine scan multiple-choice answer sheets. Thus, hand scoring of essay answers tends to increase the time required to report results. There also are concerns about subjectivity in grading because human readers do not always agree with each other (or even with themselves over time) in the score they assign to an answer. Scoring costs and logistical problems (such as arranging for readers) are much greater with open-ended tests than they are with multiple-choice exams. In addition, score reliability per hour of testing time is generally greater with multiple-choice tests than it is with open-ended ones (Wainer and Thisssen [19] and Klein and Bolus [5]). Nevertheless, there are important skills that can only be assessed (or assessed well) with open-ended measures. This is especially so in higher education. Consequently, college and graduate school admissions tests as well as licensing exams for teachers and other professionals are now likely to contain constructed response questions.

Many educators and practitioners prefer open-ended questions whereas those who administer large-scale testing programs are responsible for reducing costs and the time needed to report results. The tension between these competing interests stimulated the search for effective ways to economically and quickly score free response answers. Efforts to employ machine scoring for this purpose began about 40 years ago (Daigon [2] and Page [13]). Significant advances in this technology (particularly in computational linguistics) started to appear in the literature in the middle 1990's (e.g., Burstein et al. [1]). Machine scoring of essay answers in operational programs began a few years later (see Kukich [9] for a review of this history).

At least three private companies – Educational Testing Service (ETS), Knowledge Analysis Technologies (KAT), and Vantage Learning – provide machine grading services for a wide range of clients, such as the Army's Officer Training and Doctrine Command, industry training courses, statewide K-12 testing programs, and the National Board of Medical Examiners (Swygert et al. [18]). For example, machines are used to grade about 350,000 answers a year to the Analytic Writing Assessment section of the Graduate Management Admission Test (GMAT). Together, the three companies listed above score well over a million free response answers per year.

The computer-based methods these systems use are complex, statistically sophisticated, and proprietary. And, the results with them have been very encouraging. For example, Laudauer et al. [10] found a 0.86 correlation between two independent readers on over 2,000 essays across 15 diverse substantive topics from different grade levels. The correlation between the scores assigned by a single reader to these answers and those generated by KAT's latent semantic analysis based Intelligent Essay Assessor engine was 0.85. Similarly, Powers et al. [14] reported that in a sample of about 1800 essay answers to 40 GRE writing prompts that were graded on a six-point scale, pairs of human readers agreed either exactly or within one point of each other 99 percent of the time in the score they assigned. The corresponding agreement rate between the hand scores and those assigned by ETS's "e-rater" machine grading system was 93 percent. In short, it appears that at least in some applications, machine-assigned grades can come very close to those assigned by hand.

If such encouraging results are replicable, then machine scoring of open-ended responses could be a cost-effective way to include constructed response questions in several types of large-scale testing programs. For example, universities as well as state and community college systems could use machine scoring for placement tests and assessment programs that emphasize students demonstrating competencies rather than simply satisfying "seat-time" or credit-hour requirements. Machine scoring also is compatible with the needs of distance learning courses in that students can take the tests on line and have their scores returned quickly. Some machine scoring algorithms also provide diagnostic data regarding different aspects of answer quality (such as content, style, and mechanics).

## 1. Purposes

It is within this broader assessment context that we examined the correspondence between the hand and machine scores that were assigned to the answers to three types of college-level open-ended tasks (the issue and argument prompts that are now used on the GREs and a 90-minute critical thinking performance task). These analyses investigated the questions below to explore whether machine-assigned scores can replace hand-assigned scores in certain types of large-scale higher education assessment programs. Our specific focus was on testing conducted for research and policy analysis purposes; i.e., as distinct from college admissions, licensing, or other high-stakes applications.

- Do the hand and machine-assigned scores agree with each other as much as two hand readers agree with each other (and is this agreement rate high enough to trust the scores)?
- Are agreement rates a function of task type? For example, is the degree of agreement between hand and machine-assigned scores on a 30 or 45-minute essay question as high as it is on a 90-minute performance task in which test takers use multiple reference documents?
- Do hand and machine-assigned scores to the answers on this performance task (and to those on more typical open-ended prompts) have similar correlations with other measures (such as SAT scores and college grades)?
- How are these relationships affected when the college (rather than the individual student) is used as the unit of analysis?
- Do machine-assigned scores to one task correlate higher with machine-assigned scores to another task than they do with the hand scores on that other task? In other words, is there a unique effect related to scoring method that generalizes across tasks? This would occur if some students tended to earn higher grades when their answers were hand rather than machine scored and the reverse was true for other students.
- Does machine scoring tend to widen or narrow differences in mean scores between gender or racial/ethnic groups?
- Under what conditions is machine scoring a cost-effective strategy?

## 2. Procedures

### 2.1. Sample characteristics

The 1359 students who participated in this research were drawn from 14 colleges and universities that varied in size, selectivity, geographic location, type of funding, and the diversity of their students' background characteristics. Forty-two percent of the students were males. The percentage of freshmen, sophomores, juniors, and seniors were 29, 25, 23, and 23, respectively. Whites, African Americans, Asians, and Hispanics comprised 71, 10, 6, and 3 percent of the students, respectively (2 percent belonged to other groups, 7 percent were multiethnic, and 1 percent did not respond). Students were recruited across a broad spectrum of academic majors and were paid $20 to $25 per hour for their time (the amount varied slightly as a function of local college practices).

### 2.2. Measures

**Critical thinking tasks.** We used four of the 90-minute "Tasks in Critical Thinking" that were developed by the New Jersey Department of Higher Education (Ewell [4] and Erwin and Sebrell [3]). All of these tasks required working with various documents to answer 5 to 10 separately scored open-ended questions. Students had to interpret and critically evaluate the information in each document.

> Students are given a memo from a company officer that describes an accident in which a teenager suffered serious injury while wearing a pair of the company's high performance skates. The memo then asks the examinee to address several questions about this incident, such as the most likely reasons for it, the evidence that would support these hypotheses, and the validity of the claim that more than half of the serious skating accidents involve the company's skates. Students also receive a set of documents that they are advised to consider in preparing their memo. These materials include a newspaper account of the accident, information about the company and its market share, an accident report prepared by a custodian, transcription of an interview with an expert, and a story board for the company's TV ad. The scoring rubric assigns points for recognizing that the relationship between a company's share of the accidents and its market share and for identifying plausible reasons for the accident and the evidence to support these hypotheses. There also are overall scores for analytic reasoning and communication skills.

Fig 1. *SportsCo task.*

We also administered two new 90-minute critical thinking tasks that were developed for this study, but problems with one of them precluded using it in the analyses below. Figure 1 describes the other new task ("SportsCo"). The new tasks are similar to the New Jersey ones in that students work with various documents to prepare their answers, but unlike the New Jersey tasks, some of these materials are more germane than others to the issues that need to be addressed. Thus, students must decide how much to rely on them in preparing their answers (this corresponds to the "application of strategic knowledge" discussed by Shavelson and Huang [17]). The new tasks also differed from the New Jersey ones in that students wrote a single long answer (in the form of a memo that addressed several issues) rather than respond to a set of separate questions.

**GRE analytical writing prompts.** The Graduate Record Examination (GRE) now contains two types of essay questions, a 45-minute "issue" task and a 30-minute "argument" task (Powers et al. [14]).

The 45-minute "issue" prompt presents students with a point-of-view about a topic of general interest and asks them to respond to it from any perspective(s) they wish. One of these prompts was: "In our time, specialists of all kinds are highly overrated. We need more generalists – people who can provide broad perspectives." Students are instructed to provide relevant reasons and examples to explain and justify their views.

The 30-minute "argument" prompt presents an argument and asks test takers to critique it, discussing how well reasoned they found it, rather than simply agreeing or disagreeing with the author's position (see Figure 2 for an example).

According to Powers et al. [14]), the issue and argument "tasks are intended to complement one another. One requires test takers to construct their own arguments by making claims and providing evidence to support a position; the other requires the critique of someone else's argument" (p. 4).

**Other measures.** All the participants completed a survey that included questions about their demographic and background characteristics. They also gave us their permission to obtain their college grade point average (GPA) and SAT or ACT scores from the registrar at their college. All but one of the 14 colleges provided these scores.

> The University of Claria is generally considered one of the best universities in the world because of its instructors' reputation, which is based primarily on the extensive research and publishing record of certain faculty members. In addition, several faculty members are internationally renowned as leaders in their fields. For example, many of the faculty from the English department are regularly invited to teach at universities in other countries. Furthermore, two recent graduates of the physics department have gone on to become candidates for the Nobel Prize in Physics. And 75 percent of the students are able to find employment after graduating. Therefore, because of the reputation of its faculty, the University of Claria should be the obvious choice for anyone seeking a quality education.

Fig 2. *Example of a 30-minute GRE "argument" prompt.*

### 2.3. Test administration

At eight of the 14 colleges, each student took two of the six 90-minute critical thinking tasks. Students were assigned randomly to pairs of these tasks within schools. All six tasks were administered at each school. At the other six colleges, students were assigned randomly to one of the six 90-minute critical thinking tasks. The students were then assigned randomly to two of the three GRE issue prompts and, in keeping with the GRE's administration procedures, they were instructed to select one to answer. After completing this prompt, a student was assigned randomly to one of the three argument prompts. Thus, examinees had some choice of issue prompts but not of argument prompts. All the critical thinking tasks and GRE prompts were administered at each of these six colleges (see Klein et al. [7] for a more detailed description of the matrix sampling plan).

The critical thinking tasks and GRE prompts were presented to students in hard copy. For the critical thinking tasks, students had the option of handwriting their answers directly in their test booklets or preparing them on a computer. For the GRE prompts, students had to prepare their answers on a computer. The tests were administered to students under standardized proctored conditions usually in their college's computer lab.

### 2.4. Hand scoring SportsCo answers

Four graduate students in English from a nationally prominent university were trained to use an analytic scoring guide to evaluate the SportsCo answers. This guide contained 40 separate items (graded 0 or 1) and a 5-point overall communication score. For the latter score, the readers were told to consider whether the answer was well organized, whether it communicated clearly, whether arguments and conclusions were supported with specific reference to the documents provided, and whether the answer used appropriate vocabulary, language, and sentence structure. Readers were instructed to ignore spelling. A student's total raw score was the sum of the 41 scores.

Two of the four readers were picked at random to grade each answer. As a result of this process, every reader was paired about the same number of times with every other reader. Answers prepared on the computer were printed out so that readers evaluated hard copies of all the answers. The reader-assigned scores are hereinafter referred to as the "hand" raw scores so as to distinguish them from the machine-assigned scores.

| Reader | Mean raw score | Standard deviation | Mean correlation with other readers |
|--------|----------------|--------------------|-------------------------------------|
| 1 | 11.69 | 4.80 | 0.86 |
| 2 | 11.29 | 3.86 | 0.83 |
| 3 | 11.35 | 3.88 | 0.85 |
| 4 | 12.23 | 4.34 | 0.85 |
| Mean | 11.60 | 4.15 | 0.85 |

Table 1 shows each reader's mean and standard deviation as well as each reader's mean correlation with the other readers. The overall average correlation between two readers was 0.85. The small differences in means between readers were not significant.

### 2.5. Machine scoring SportsCo answers

We gave ETS a sample of 323 SportsCo answers and the individual item scores assigned to them by each reader. ETS used these data to construct its machine-grading algorithms. To build the e-rater algorithm for the communication score, ETS randomly divided the 323 students into three sets (labeled *A*, *B*, and *C*). Next, it used the answers and scores for the students in Sets *A* and *B* to build a model that was then applied to the answers in Set *C*, it used Sets *A* and *C* to build a model that was then applied to the answers in Set *B*, and it used Sets *B* and *C* to build a model that was then applied to the answers in Set *A*. The machine-assigned scores created by this process were therefore independent of the scores that were used to build the machine scoring models.

ETS's "e-rater" scoring engine "is designed to identify features in student essay writing that reflect characteristics that are specified in reader scoring guides" (Burstein, et al. [1]). It develops a scoring algorithm that is based on the grades assigned by the human readers to a sample of answers and contains modules for identifying the following features that are relevant to the scoring guide criteria: syntax, discourse, topical content, and lexical complexity.

ETS's "c-rater" scoring system (Leacock and Chodorow [11]) was used to create scores for items 1 through 40. This engine is designed for content-laden short answer questions (rather than for evaluating the long memo examinees had to produce for this task). Thus, it had to locate that portion in the student's answer that was relevant to each item in the scoring rubric as well as grade it. The models that were relatively easy to build typically had a high agreement between readers, a low "baseline" (i.e., relatively few students received credit for the item), and had a well-defined rubric with two to three main ideas that receive credit. For example, on item #4, students received credit for recognizing that SportsCo had double (or "more") sales than its chief competitor (the AXM Company). There were two generically correct responses to this item, namely: (1) SportsCo sold twice as many skates as AXM and (2) AXM sold half as many skates as SportsCo.

There are many ways in which a student can express these ideas. Hence, the model included several variations of each sentence. By utilizing synonyms and selecting the essential features, the model can identify a variety of paraphrases. For example, the first sentence matches the student who wrote, "The reason for this is most likely because we also sell more than twice as many skates than any other skating manufacturer," and the student who wrote, "We had twice as many injuries due to selling twice as many skates."

Low agreement between hand and c-rater scores can occur as a result of several factors. For example, item #35 gave students credit for saying "the warning was not strong enough." Instead of saying this, many students offered a solution, such as by describing how the warning should be revised. Although it was clear that the intent of these suggestions was to make the warning stronger, c-rater could not recognize this intent.

The training and cross-validation sets for the c-rater data were not created in the same manner as those for e-rater. For c-rater, the scored responses for each prompt were partitioned individually. About one-third of the responses were used for training (to be consulted while manually building the c-rater models), and the rest for blind cross validation. However, a maximum of 50 responses were kept for each item in the development set. Thus, if there were 168 responses receiving credit, 50 went to training and 118 to cross-validation. In addition, if there were 15 or fewer responses receiving credit in the entire dataset, these were evenly divided between training and cross-validation.

Finally, the degree of agreement between hand readers sets an upper limit on the degree of agreement between the hand and machine-assigned scores. This is true for both the "e-rater" and "c-rater" scoring engines.

### 2.6. Hand and machine scoring of answers to other tasks

A two-person team hand graded the answers to the four critical thinking tasks that were developed by New Jersey and to the six GRE prompts. This team had extensive prior experience in scoring responses to these tasks. The responses to a task were assigned randomly to readers. To assess reader agreement, both readers also independently graded a common batch of about 15 percent of the answers to each task.

On the four New Jersey critical thinking tasks (Conland, Mosquitoes, Myths, and Women's Lives), the graders used a blend of analytic and holistic scoring rubrics depending on the particular question being scored. The mean correlation between readers on the total scores on these tasks ranged from 0.88 to 0.93 with a mean of 0.90 (which is slightly higher than the 0.85 on SportsCo).

The readers graded the answers to the GRE prompts on a holistic six-point scale; i.e., they read the answers quickly for a total impression that considered syntactic variety, use of grammar, mechanics, and style, organization and development, and vocabulary usage. The mean correlation between two readers on a 45-minute issue prompt and a 30-minute argument prompt were 0.84 and 0.86, respectively.

The "e-rater" scoring engine was used to evaluate the answers to the GRE prompts using algorithms that were developed previously (Schaeffer et al. [16]). Thus, it was not necessary to implement the model building process that was required for SportsCo.

### 2.7. Scaling

We used a standard conversion table to put ACT scores on the same scale of measurement as SAT scores. The converted scores are hereinafter referred to as SAT scores. We transformed the GPAs within a college to z-scores. To adjust for possible differences in grading standards across colleges, we also scaled the GPAs within a college to a score distribution that had the same mean and standard deviation as its students' SAT scores (these are hereinafter referred to as "adjusted GPAs").

As noted above, we used three GRE issue prompts and three GRE argument prompts. To adjust for possible differences in difficulty among these prompts and to facilitate combining scores across prompts, the reader assigned "raw" scores on a prompt were converted to a score distribution that had the same mean and standard deviation as the SAT scores of the students who took that prompt. We did the same thing with the machine-assigned scores on each prompt and with the scores on the New Jersey tasks.

## 3. Analyses and results

### 3.1. Relationship between hand and machine assigned scores

There were 323 students who had both hand and machine-assigned scores on their answers to SportsCo, 590 students who had hand and machine-assigned scores on at least one GRE issue prompt and one argument prompt, and 79 students who were in both of these two groups. The correlation between two hand readers was usually only slightly higher than it was between the hand and computer assigned scores (Table 2). The one exception was on the GRE argument prompt where the correlation between two hand readers was 0.19 higher than it was between the hand and machine scores.

The hand readers tended to assign slightly higher raw scores to SportsCo items 1-40 than did the c-rater algorithm whereas they assigned slightly lower raw communication scores than did the e-rater algorithm (Table 3). There was almost no difference between the mean hand and mean machine-assigned GRE writing scores but that may be a by-product of the scaling described above. Standard deviations also were quite comparable.

There did not appear to be a systematic effect of scoring method across tasks. For instance, SportsCo *machine* scores correlated higher with GRE *hand* scores then they did with GRE machine scores (Table 4). Moreover, the correlation between GRE issue and argument prompts was not consistently higher when the same scoring method was used then when different methods were used (Table 5).

Table 2
*Correlation between two hand readers and between hand and machine assigned scores by task*

| Task | Score | Between two hand readers | Between hand and machine scores |
|------|-------|--------------------------|----------------------------------|
| SportsCo | Sum of items 1 to 40 | 0.84 | 0.81 |
| | Communication (41) | 0.61 | 0.57 |
| | Total (1 to 41) | 0.85 | 0.83 |
| GRE | Issue Prompt | 0.84 | 0.73 |
| | Argument Prompt | 0.86 | 0.67 |
| | Mean GRE | | 0.78 |

Table 3
*Means and standard deviations by scoring method and task*

| Task | Score type | Mean | | Standard deviation | |
|------|-----------|------|---------|--------------------|---------|
| | | Hand | Machine | Hand | Machine |
| SportsCo | Sum of items 1 to 40 | 8.71 | 7.66 | 3.47 | 3.35 |
| | Communication (41) | 2.89 | 3.15 | 0.94 | 1.10 |
| | Total Raw (1 to 41) | 11.60 | 10.80 | 4.15 | 3.97 |
| GRE | Issue Prompt | 1118 | 1126 | 188.5 | 189.1 |
| | Argument Scale | 1111 | 1110 | 183.5 | 182.4 |
| | Mean GRE Scale | 1114 | 1118 | 160.5 | 165.0 |

### 3.2. Relationship of hand and machine scores to scores on other tests

SportsCo's hand scores correlated with other indexes of academic ability to about the same degree as its machine scores correlated with those measures. For example, SportsCo hand and machine scores correlated 0.36 and 0.34, respectively, with unadjusted College GPA. Although SportsCo is only a 90-minute task, this relationship was almost as strong as the one between SAT scores and College GPA in our samples (this correlation was 0.36 among students with SportsCo scores and 0.31 among those with GRE scores). GRE hand and machine scores also had comparable relationships with other test scores (Table 6). These findings indicate that the two scoring methods yield very similar results.

### 3.3. Relationship of scoring method to student characteristics

Males and females had very similar mean scores on all the measures regardless of scoring method. For example, the mean hand and machine total GRE scores for females were 1112 and 1122, respectively; i.e., a difference of 10 points or about 0.06 standard deviation units. The means for males were almost identical, 1121 and 1118, respectively.

TABLE 4
*Correlation between total SportsCo and GRE total scores (N = 79)*

|  | SportsCo hand | SportsCo machine |
|---|---|---|
| GRE Hand Score | 0.53 | 0.58 |
| GRE Machine Score | 0.36 | 0.43 |

TABLE 5
*Correlation between GRE prompt types when the same versus different scoring methods are used*

| Argument | Issue | Correlation |
|---|---|---|
| Hand | Hand | 0.49 |
| Machine | Machine | 0.58 |
| Hand | Machine | 0.45 |
| Machine | Hand | 0.55 |

TABLE 6
*Correlation of hand and machine scores with scores on other measures*

| Score | SportsCo | | GRE Total | |
|---|---|---|---|---|
| | Hand | Machine | Hand | Machine |
| SAT total | 0.51 | 0.40 | 0.61 | 0.55 |
| College GPA | 0.36 | 0.34 | 0.25 | 0.22 |
| Adjusted College GPA | 0.53 | 0.48 | 0.52 | 0.48 |
| NJ Women's Lives | 0.50 | 0.43 | 0.56 | 0.63 |
| NJ Mosquitos | 0.50 | 0.49 | 0.60 | 0.60 |
| NJ Conland | | | 0.54 | 0.58 |
| NJ Myths | | | 0.57 | 0.53 |

Notes: SportsCo was paired with only two of the four New Jersey critical thinking tasks. There were about 90 students who took a given NJ task and also had both a hand and machine GRE total score. The same was true for SportsCo.

As a group, the combination of whites and Asians had significantly higher mean scores on all the measures than did other students, but the size of this difference was similar across scoring methods. For example, the White/Asian hand-scored mean on SportsCo was 12.25 whereas it was 9.64 for all other students combined (i.e., a difference of 2.61 points or about two-thirds of a standard deviation). The mean machine scores of these two groups were 11.35 and 9.16, respectively (i.e., a difference of 2.19 points). Thus, the net effect of scoring method on the difference between these two groups was less than one half of one point $(2.61 - 2.19 = 0.42)$ or about one-tenth of a standard deviation.

There were similar results on the GRE. As a group, Whites and Asians had a mean of 1147 when their answers were hand scored and a mean of 1146 when they were machine scored. The corresponding means for all other students combined were 1025 and 1041 (i.e., a 16 point difference between scoring methods which is again about one-tenth of a standard deviation). These data suggest that machine scoring is more likely to ameliorate than exacerbate the differences in mean scores between racial/ethnic groups.

### 3.4. Unit of analysis effects

Certain types of research and policy analysis studies use the school rather than the student as the unit of analysis (Meyer [12]). These investigations examine differences in mean scores among schools (or other aggregations of students) such as to assess whether the students at a college generally score higher or lower than would be expected given their SAT scores (see Klein et al. [7] for an example of this type of study).

When the college is the unit, the correlation between the total GRE hand and machine-assigned scores was 0.99. The school level correlation between hand and machine scores was 0.95 on SportsCo (compared to 0.85 when the student is the unit). Using the college as the unit also increases the correlation among other tests. This occurs with both the hand and machine scores. For example, when the student is the unit, the hand scores on a New Jersey critical thinking task correlate 0.50 and 0.46 with the hand and machine scores on SportsCo, respectively. When the college is the unit, the corresponding correlations are 0.91 and 0.86. These similarly high correlations suggest that the machine scores alone can be relied on for studies that use the college as the unit. Hand scores would almost be redundant.

### 3.5. Cost effectiveness

Assessment of learning in higher education has become a critical concern to policy makers, college accreditation agencies, educators, students, and parents. Yet to do so without submitting to the serious limitations of multiple choice tests is expensive in part because of the costs of scoring the answers to open-ended questions. Hence, one of the purposes of this study was to examine the utility of using computer scoring of different types of constructed response questions to see if there was now a more efficient way (in terms of costs and scoring time) to help solve this part of the assessment puzzle.

We found that the costs of machine scoring depend on several factors. First, all the answers have to be in machine readable form. Realistically, this means that students have to key enter their responses, such as by taking the test at one of their college's computer labs. Because of limited space and computer availability, this requirement may lead to administering the measures over several days when a hundred or more students are tested at a college (test security concerns may preclude allowing students to take the tests in their homes or dorm rooms). Efficient procedures are needed to send the students' responses to the scoring service, such as by uploading them over the Web.

Second, machine-scoring costs are a function of the number of questions and the number of answers to them. For instance, one company advertised (in 2003) a setup fee of $75,000 to develop the scoring algorithms for the first ten prompts and $2,500 for each additional prompt. It also advertised a charge of 85-cents per answer. Under these terms, the direct cost of machine scoring (including the cost of hand scoring a sample of 250 answers that the computer needs to "learn" on) is less than that of hand scoring if (1) there are over 4,000 answers to be scored to each of 16 different prompts and (2) it costs $2.50 or more to hand-score an answer once.

There are, however, several significant indirect costs that also have to be considered. For example, hand scoring often requires training large numbers of readers and the handling, shipping, and keeping track of numerous boxes of answers. This can be a staff intensive, time consuming, and complex logistical process that requires lots of staff and space.

Third, machine scoring (which takes less than eight seconds per answer) is much faster than hand scoring (and this can be done by multiple computers working simultaneously that do not need coffee or rest breaks). Scoring time is an important consideration for programs that have to report results promptly. In addition, machine scoring may be better suited to testing programs that are conducted over several months because unlike hand scoring, it does not require constantly arranging for readers to score relatively small batches of answers.

Because of these and other factors, the Council for Aid to Education's Collegiate Learning Assessment (CLA) program delivers tests to students electronically over the Internet. Examinees answer online and their responses are uploaded for machine scoring. The per student cost of this system is less than one-fourth of that required for the hardcopy test booklet and hand scoring procedures that were used in the study above (for details, see [www.cae.org](www.cae.org)).

## 4. Discussion and conclusions

We examined the correspondence between the hand and machine scores that were assigned to the answers to three types of college-level open-ended tasks (a 90-minute performance task and the issue and argument prompts that are now used on the GREs). These analyses found that the machine-assigned scores could replace the hand assigned scores for at least some types of large-scale testing programs (such as for studies that assessed whether a college's students are generally scoring higher or lower than would be expected on the basis of their SAT scores). The specific findings that support this conclusion are as follows:

- Hand and machine-assigned scores agreed highly with each other. And, the degree of agreement between these scoring methods was about as high as it was between two hand readers. This was true on SportsCo (the 90-minute performance task) and on the GRE prompts. More recent analyses of these measures with much larger samples confirm these findings (Klein, Shavelson, Benjamin and Bolus [8]).

- There is a near perfect correlation (0.95) between the hand and machine scores on the 90-minute SportsCo performance task when the college (rather than the student) is the unit of analysis. It was 0.99 on the GRE.
- Hand and machine-assigned scores have very similar correlations with other constructed response and selected response measures, such as SAT scores, the grades on other critical thinking tasks, and college GPAs. In other words, the hand and machine-assigned scores behave the same way.
- The machine-assigned scores to one task correlate about as well with the machine-assigned scores to another task as they do with the hand scores on that other task. Thus, there is no indication that some students consistently earn higher grades when their answers are hand versus machine scored.
- There is no interaction between gender and scoring method, and if anything, machine scoring tends to very slightly narrow the differences in mean scores between racial/ethnic groups. It certainly does not widen them.
- It appears that the economic benefits of machine scoring can be realized when there are a few thousand or more answers to be scored.

The results above were obtained in a "low-stakes" research study where students were paid to participate. It is not certain whether comparable results would be obtained with a different set of open-ended measures or if there were significant external incentives for students to "beat" the machine (Powers et al. [15]), such as in a college admissions or licensing context. Nevertheless, the research described above used a variety of free response tasks and the students who participated in it generally reported that these tasks were engaging and they tried to do their best on them (Klein et al. [6]). The substantial correlations of their scores with SAT scores and college GPAs certainly suggest they took the measures seriously.

Regardless of scoring method, there was about a two-thirds of a standard deviation difference on our open-ended measures between the mean scores of the two clusters of racial/ethnic groups studied (i.e., Whites plus Asians versus all others) whereas there was over a one full standard deviation difference between these groups on the SAT. For example, their mean total SAT scores in the GRE sample were 1162 and 968, respectively. While some of this disparity is no doubt due to the SAT scores having a higher reliability than the free response measures, it nevertheless suggests that the kinds of open-ended tasks used in this research might tend to slightly narrow the gap in mean scores among groups.

Finally, it appeared that the study's open-ended tasks captured important abilities that are not fully assessed by more traditional multiple-choice measures. For instance, college GPAs correlated about as well with the scores on our critical thinking measures as they did with SAT scores (the median of the within school correlations of college GPA with the scores on our critical thinking tasks and with SAT scores were 0.33 and 0.36, respectively). However, at over half of the participating colleges, combining SAT scores with the scores on open-ended measures yielded a statistically significantly better prediction of college GPAs than did SAT scores alone (at $p < 0.05$).

Taken together, these findings suggest that adding the kinds of tests used in this research to a battery that already contains SAT or ACT scores may improve overall predictive validity while at the same time slightly narrow differences in mean scores between racial/ethnic groups. Machine scoring the answers to these or other types of open-ended tasks also may be key to making their inclusion a practical option. It will be up to future studies to examine this matter as well as the utility of machine scoring for high-stakes testing programs.